\documentclass[sigconf]{acmart}
\usepackage{graphicx}
\usepackage{subcaption}
\usepackage[linesnumbered,ruled,vlined,noend]{algorithm2e}
\usepackage{multirow}
\usepackage{enumerate}
\usepackage{adjustbox}
\graphicspath{{images/}}

\definecolor{comment-color}{rgb}{0.216, 0.455, 0.424}
\AtBeginDocument{%
  \providecommand\BibTeX{{%
    \normalfont B\kern-0.5em{\scshape i\kern-0.25em b}\kern-0.8em\TeX}}}

\copyrightyear{2021}
\acmYear{2021}
\setcopyright{acmlicensed}\acmConference[MM '21]{Proceedings of the 29th ACM International Conference on Multimedia}{October 20--24, 2021}{Virtual Event, China}
\acmBooktitle{Proceedings of the 29th ACM International Conference on Multimedia (MM '21), October 20--24, 2021, Virtual Event, China}
\acmPrice{15.00}
\acmDOI{10.1145/3474085.3475182}
\acmISBN{978-1-4503-8651-7/21/10}

\acmSubmissionID{mfp0146}

\begin{document}
\fancyhead{}
\title{Joint Optimization in Edge-Cloud Continuum for Federated Unsupervised Person Re-identification}






\author{Weiming Zhuang$^{1, 3}$ \quad Yonggang Wen$^{2}$ \quad \quad Shuai Zhang$^{3}$}
\affiliation{%
 $^{1}$S-Lab, Nanyang Technological University \enspace $^{2}$Nanyang Technological University \enspace $^{3}$SenseTime Research
}
\email{weiming001@e.ntu.edu.sg, ygwen@ntu.edu.sg, zhangshuai@sensetime.com}

\renewcommand{\authors}{Weiming Zhuang, Yonggang Wen, and Shuai Zhang}

\renewcommand{\shortauthors}{W. Zhuang et al.}

\begin{abstract}

Person re-identification (ReID) aims to re-identify a person from non-overlapping camera views. Since person ReID data contains sensitive personal information, researchers have adopted federated learning, an emerging distributed training method, to mitigate the privacy leakage risks. However, existing studies rely on data labels that are laborious and time-consuming to obtain. We present \textit{FedUReID}, a federated unsupervised person ReID system to learn person ReID models without any labels while preserving privacy. FedUReID enables in-situ model training on edges with unlabeled data. A cloud server aggregates models from edges instead of centralizing raw data to preserve data privacy. Moreover, to tackle the problem that edges vary in data volumes and distributions, we personalize training in edges with joint optimization of cloud and edge. Specifically, we propose \textit{personalized epoch} to reassign computation throughout training, \textit{personalized clustering} to iteratively predict suitable labels for unlabeled data, and \textit{personalized update} to adapt the server aggregated model to each edge. Extensive experiments on eight person ReID datasets demonstrate that FedUReID not only achieves higher accuracy but also reduces computation cost by 29\%. Our FedUReID system with the joint optimization will shed light on implementing federated learning to more multimedia tasks without data labels.

\end{abstract}

\ccsdesc[500]{Computing methodologies~Distributed algorithms}
\ccsdesc[500]{Information systems~Top-k retrieval in databases}
\ccsdesc[300]{Computing methodologies~Object identification}
\ccsdesc[300]{Computing methodologies~Matching}

\keywords{federated learning, person re-identification, unsupervised learning, unsupervised person re-identification}

\maketitle


\section{Introduction}

Person re-identification (ReID) is an important computer vision task that has attracted considerable research interests in the multimedia community due to its wide applications on public safety and public security \cite{leng2019survey}. Person ReID aims to match the same person in non-overlapping camera views. Over the years, researchers have proposed many innovative approaches, based on either supervised learning or unsupervised learning, to excel the performance of person ReID \cite{ye2021deep-survey, lin2019buc, zeng2020hierarchical-hct}. The majority of these methods require centralizing plenty of images into a central server.

However, the increasingly stringent data privacy regulations limit the development of person ReID \cite{gdpr}. The data collected from cameras contain sensitive personal information such as the identity and location of individuals. Due to data privacy concerns, centralizing the data would not be feasible because it would impose potential privacy leakage risks.

A recent study \cite{zhuang2020fedreid} proposes federated person re-identification (FedReID) to train person ReID models while preserving data privacy. Federated learning (FL) is a distributed learning technique that allows multiple parties to train models collectively without centralizing data \cite{fedavg}. FedReID implements FL to person ReID. It achieves outstanding performance and effectively preserves data privacy by transmitting model updates instead of centralizing data. However, FedReID heavily relies on the assumption that data has labels in clients. In real-world applications, annotating data is expensive, laborious, and time-consuming. Relying on labels is also not scalable for large-scale deployment. It takes three annotators two months to produce 126,441 bounding boxes of 4,101 identities from 180 hours of videos \cite{Wei2017Msmt}.

In this paper, we present a new federated unsupervised person ReID system, \textit{FedUReID}, to learn person ReID models from unlabeled data while preserving data privacy. FedUReID stores data collected from multiple cameras in edges. Edges are coordinated by a cloud server to train models with unsupervised person ReID methods like Bottom-Up Clustering (BUC) \cite{lin2019buc}. FedUReID preserves data privacy because edges only access their local raw data. However, the training in edges suffers from statistical heterogeneity. Since data are collected from cameras in various locations, they would vary in the number of images, the number of identities, and data distributions affected by the data collection environment. These variances cause \textit{statistical heterogeneity} \cite{Li2020FedChallenges, zhuang2020fedreid} among edges, affecting the performance of the system. 


To tackle the statistical heterogeneity among edges, we propose joint optimization of cloud and edge to personalize models for each edge. In particular, we introduce three optimizations, \textit{personalized epoch (PE)} and \textit{personalized clustering (PC)} in the edges and \textit{personalized update (PU)} in the cloud. (1) PE: unlike traditional FL that trains the same epochs in all training rounds, PE dynamically adjusts the number of training epochs by early stopping a training round when conditions are satisfied. (2) PC: for unlabeled data, edges adopt a hierarchical clustering method to iteratively predict labels by merging similar clusters in each training round. It regards each data point as a cluster at the start. Unlike simple implementation that all edges merge the same percentage of clusters, PC customizes the merge percent for edges such that the numbers of final clusters are similar to their actual labeled number of identities. (3) PU: the cloud server aggregates model updates from edges to obtain a new model. Unlike traditional FL that directly updates models in edges with the aggregated model, PU updates these models with exponential moving average (EMA) to better adapt the aggregated model to each edge. We calculate the weighting of EMA using the similarity between the cloud aggregated model and the edge local models, measured by normalized Euclidean distance. 

Extensive experiments and ablations on eight datasets demonstrate the effectiveness of FedUReID with joint optimization. We first construct the baseline with a simple combination of FL \cite{zhuang2020fedreid} and unsupervised person ReID training \cite{lin2019buc}. It does not outperform standalone training in all datasets. Standalone training means that each edge conducts unsupervised person ReID training \cite{lin2019buc} with its local dataset. In contrast, any single optimization method (PE, PC, or PU) outperforms standalone training and the baseline. Joint optimization of all optimization methods achieves the best performance. Compared with standalone training, it improves rank-1 accuracy by over 18\% on the two smallest datasets \cite{3dpes, iLIDS-VID}. Moreover, compared with the baseline, it not only improves 4\% on the two largest datasets \cite{zheng2017dukemtmc-reid, Zheng2015Market1501}, but also reduces computation cost by 29\%.


In this paper, we make the following contributions: 

\begin{itemize}
    \item We propose the \textit{first} federated unsupervised person ReID system. It learns person ReID models without any labels while preserving data privacy.
    
    \item We propose joint optimization of cloud and edge to address the statistical heterogeneity among edges via personalization. In particular, we introduce personalized epoch, personalized clustering, and personalized update. 
    
    \item We demonstrate the effectiveness of our proposed optimizations via extensive experiments and ablation studies.
\end{itemize}

The rest of the paper is organized as follows. In Section \ref{sec:related-work}, we review related work about unsupervised person ReID and federated learning. Section \ref{sec:method} introduces our proposed FedUReID with joint optimization of cloud and edge. We present the experimental results and analyze the optimization methods via ablations in Section \ref{sec:experiments}. In Section \ref{sec:conclusion}, we summarize the paper and provides future directions.

\section{Related Work}
\label{sec:related-work}

\subsection{Unsupervised Person Re-identification}

Person ReID aims to match a person in non-overlapping camera views. Supervised person ReID has achieved outstanding performance over the years of development \cite{zheng2016person-reid-survey, ye2021deep-survey, wang2018learning, liu2020domain, lan2020magnifiernet}. Recently, unsupervised person ReID is receiving increased attention \cite{fan2018unsupervised-tomm-pul, zhong2018generalizing-hhl, lin2019buc, zeng2020hierarchical-hct}. Most unsupervised person ReID methods fall into two categories: unsupervised domain adaptation (UDA) and purely unsupervised. 

\textbf{Unsupervised Domain Adaptation for Person ReID} UDA aims to learn a model that performs well for unlabeled data in a target domain, given labeled data in a source domain. These two domains differ in data distributions. On the one hand, some studies improve target domain performance by transferring image styles from the source to target domain \cite{wei2018person-gan, deng2018spgan, liu2019adaptive-ATNet} based on generative adversarial networks (GAN) \cite{goodfellow2014GAN}. On the other hand, some studies use clustering-based approach \cite{fan2018unsupervised-tomm-pul, zhuang2021fedfr} or graph matching \cite{ye2017dynamic-graph} to generate pseudo labels for the unlabeled data. Besides, HHL \cite{zhong2018generalizing-hhl} leverages camera invariance and domain connectedness to obtain a generalized model for the target domain. However, all these methods would impose potential privacy leakage because they require co-locating data from both domains.

\textbf{Purely Unsupervised Person ReID} Unlike UDA methods that assume some data has labels, purely unsupervised person ReID does not rely on any labels, which is even more challenging. Researchers mainly leverage bottom-up clustering methods \cite{lin2019buc, Chen-content-analysis} to predict pseudo labels for the unlabeled data. These methods iteratively generate new pseudo labels and update the classifiers. However, these methods are not satisfactory for datasets with small data volumes. They require centralizing a large amount of data, which imposes potential privacy leakage risks. In this paper, built on the bottom-up clustering (BUC \cite{lin2019buc}) method for unsupervised person ReID training in each client, we propose joint optimization of cloud and edge to elevate performance while preserving data privacy.



\subsection{Federated Learning}

Federated learning (FL) is an emerging technique for training with decentralized data without privacy leakage \cite{fedavg}. FL trains models collectively from distributed clients under the collaboration of a central server. Federated person re-identification (FedReID) implements federated learning to person ReID \cite{zhuang2020fedreid}. It proposes Federated Partial Averaging (FedPav) to aggregate part of the models from clients. We integrate FedPav with BUC as the baseline.

\textbf{Unsupervised Federated Learning} The majority of studies on FL are based on supervised learning \cite{fedavg, fedma, zhuang2020fedreid}. Recently, several studies investigate unsupervised FL \cite{van2020fedae, zhang2020fedca}. But these methods are inapplicable to person ReID because they mainly focus on learning generic representations.

\textbf{Statistical Heterogeneity in FL} Statistical heterogeneity is one of the key challenges of FL \cite{Li2020FedChallenges, kairouz2019advances}. It has attracted extensive research interests in recent years \cite{zhao2018non-iid, xin2019meta, fedprox, zhuang2020fedreid}. Among them, some studies propose to personalize models for clients. These personalized federated learning methods are based on meta-learning \cite{dinh2020personalized-meta, fallah2020personalized-meta}, multi-task learning \cite{smith2017personalized-multitask}, knowledge distillation \cite{li2019fedmd, zhuang2020fedreid}, etc. However, these methods assume that data has labels. Our proposed FedUReID introduces joint optimization of cloud and edge to personalize clients without any labels.

\begin{figure}[h]
\begin{center}
\includegraphics[width=\linewidth]{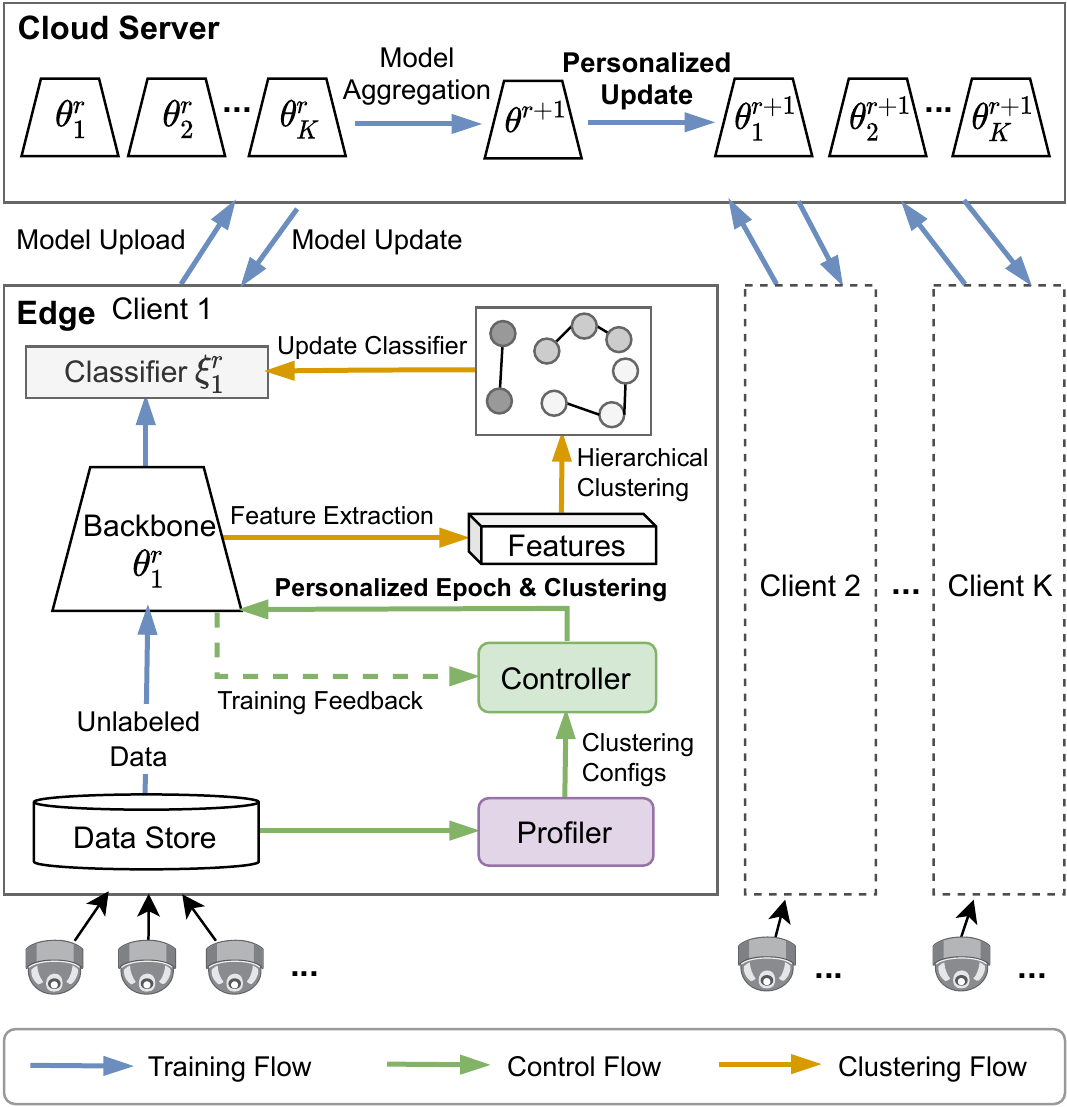}
\end{center}
    \caption{Architecture of our federated unsupervised person ReID system, \textit{FedUReID}. FedUReID consists of three flows: training flow to learn models, control flow to control training, and clustering flow to predict pseudo labels for unlabeled data. We propose joint optimization of cloud and edge: \textit{personalized epoch} and \textit{personalized clustering} in the edges, and \textit{personalized update} in the cloud. }
\label{fig:architecture}
\end{figure}

\section{Methodology}
\label{sec:method}

In this section, we present \textit{FedUReID}, a new federated unsupervised person ReID system with joint optimization of cloud and edge, to learn models without any labels while preserving privacy. It effectively tackles the statistical heterogeneity among edges.





\subsection{Overview}


We first provide an overview of FedUReID. Figure \ref{fig:architecture} depicts the system architecture of FedUReID. We embrace a hierarchical system design with a cloud server, multiple edges, and several cameras connected to each edge. Each edge is regarded as a client. FedUReID preserves data privacy because only the edge that collects data accesses it.

We design three flows for the training process: training flow, control flow, and clustering flow. At the start of the \textit{training flow}, the server initializes a model and distributes it to clients. For each training round $r$, the server and clients collaborate to train models with four steps: (1) Local training: each client $k$ trains the concatenation of the model $\theta_k^r$ and a locally initialized classifier $\xi_k^r$; (2) Model upload: each client $k$ uploads the model $\theta_k^r$ to the server; (3) Model aggregation: the server aggregates these models to obtain a new global model $\theta^{r+1}$; (4) Model update: the server updates clients' models for the next training round. \textit{Control flow} and \textit{clustering flow} interact with the training flow in the first step, local training. Before local training of the first training round, the control flow profiles the clients for clustering configurations. During local training, the control flow controls training according to training feedback. After local training, the clustering flow predicts new pseudo labels for the next round via hierarchical clustering with clustering configurations. Then, it updates the dimensions of the classifier.

Within these three flows, we propose joint optimization of cloud and edge: two optimizations in the edge --- \textit{Personalized Epoch} (PE) and \textit{Personalized Clustering} (PC), and one optimization in the cloud --- \textit{Personalized Update} (PU). These optimizations address the statistical heterogeneity among clients. We summarize FedUReID with joint optimization in Algorithm \ref{algo:fedureid}. Next, we explain these optimizations in detail.


\subsection{Client Design}
\label{sec:client-design}

Clients are responsible to perform unsupervised person ReID training. We adopt the hierarchical clustering algorithm \cite{lin2019buc} to train models with unlabeled data. The unlabeled data is collected from multiple connected cameras and stored in \textit{Data Store}. As the cameras could be deployed in various places, clients could have large variations in the number of images, the number of identities, and data distributions, leading to \textit{statistical heterogeneity} among these clients. To address it, we propose two components to personalize training in each edge: (1) a \textit{Profiler} that generates customized clustering configurations; (2) a \textit{Controller} that personalizes clustering with the configurations and reassigns computation throughout training according to training feedback.


\begin{figure}[t]
\begin{center}
\includegraphics[width=0.8\linewidth]{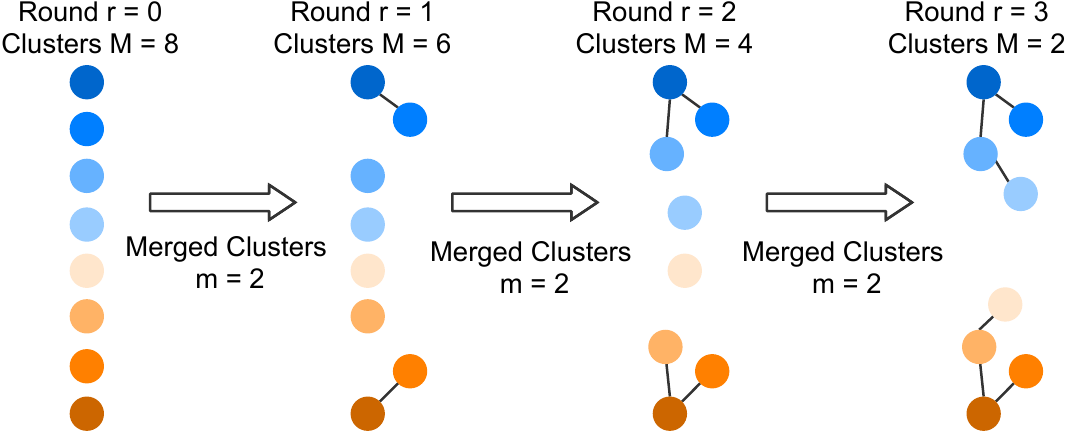}
\end{center}
    \caption{Illustration of hierarchical clustering. Each circle represents a sample. It starts with $M=8$ samples (clusters) and merges $m=2$ clusters each round.}
\label{fig:buc}
\end{figure}

\subsubsection{\textbf{Hierarchical Clustering}} 
\label{sec:baseline}
At the end of each training round, each client uses hierarchical clustering \cite{lin2019buc} to predict labels for the next training round. The clustering flow is illustrated in Figure \ref{fig:architecture}. Firstly, client $k$ extract features using the trained model $\theta^r_k$. Then, these features are merged based on similarities to form new clusters, where each cluster is regarded as a label. Lastly, the classifier is updated with the new dimension equaling to the number of clusters. Figure \ref{fig:buc} illustrates the clustering process intuitively. Client $k$ initializes the number of clusters $M$ to be the same as the number of training samples $n_k$. The number of merged clusters each round is $m = n_k * mp$, where merge percent $mp \in (0, 1)$. After clustering, the number of clusters is updated with $M = M - m$, so as the dimension of the classifier. As the number of clusters turns to 0 after $\frac{1}{mp}$ times of clustering, the maximum training round is $\frac{1}{mp}$. 

\textbf{Baseline Method and Problems} We use the direct combination of hierarchical clustering and federated learning as the \textit{baseline method}. In particular, all clients first train $E$ local epochs in all rounds, and then cluster with the same merge percent $mp$. However, due to statistical heterogeneity of clients, \textit{such combination is problematic}: (1) Using the same local epoch causes inappropriate computation assignment in different training rounds (Section \ref{sec:pe}). (2) Using the same merge percent causes inappropriate clustering paces among clients (Section \ref{sec:pc}). We analyze these two problems in detail and propose two optimization methods for them.

\SetKwInput{KwInput}{Input}                
\SetKwInput{KwOutput}{Output}              
\SetKwFunction{FnClient}{}
\SetKwFunction{FnServer}{}

\begin{algorithm}[t]
    \caption{Federated Unsupervised Person ReID}
    \label{algo:fedureid}
    \SetAlgoLined
    \KwInput{Local epoch $E$, batch size $B$, training round $R$, number of selected clients $K$, number of clients $N$, learning rate $\eta$, data size $n$, embedding size $v$, data $\mathcal{X}$}
    \KwOutput{Personalized model $\theta_k^R$ of client $k$, Global model $\theta^R$}
    
    \SetKwProg{Fn}{Server}{:}{}
    \Fn{\FnServer}{
        Initialize model $\theta^0$ and $\theta_k^0$ for each client $k$ \;
        \For{each round r = 0 to R-1}{
            $S_t \leftarrow$ (randomly select K out of N clients)\;
            \For{each client $k \in S_t$ in parallel}{
                $\theta^{r}_k \leftarrow $ \textbf{Client}($\theta_k^r$, $k$, $r$) \;
            }

            $\theta^{r+1} \leftarrow \sum_{k \in S_t} \frac{n_k}{n} \theta^{r}_k$ ; // \textcolor{comment-color}{Model aggregation} \
            
            \textcolor{comment-color}{// Personalized update}
            
            \For{each client $k \in S_t$}{ \label{line:pu-start}
                \For{each layer l = 1,2,..,L of $\theta_k^r$}{
                    $similarities_l \leftarrow \Vert \theta^{r+1,l} - \theta^{r,l}_{k} \Vert_2^2$ \; 
                }
                $\mu \leftarrow \frac{\sum_{l = 1}^L normalize(similarities_l)}{L}$ \; \label{line:pu-end}
                $\theta_k^{r+1} \leftarrow \mu \theta_k^r + (1-\mu) \theta^{r+1} $ \; 
            }
        }
        \KwRet $\theta^{R}$, $\theta_k^{R}$ of each client $k$ \;
    }
    
    \SetKwProg{Fn}{Client}{:}{\KwRet}
    \Fn{\FnClient{$\theta$, k, r}}{
        \If{r == 0}{
            Initialize the number of clusters $M \leftarrow n_k$ \;
            Initialize classifier $\xi$ with dimension $v \times M$ \;
            Initialize pseudo labels $\mathcal{Y} \leftarrow \{y_i=i\}_{i=0}^{M-1}$ \;
            $m, mp \leftarrow $ (profile client with Eqn \ref{eq:pc-1} and \ref{eq:pc-2}) ; \label{line:profiling}
        }
        
        $\mathcal{B} \leftarrow $ (split data $\{\mathcal{X, Y}\}$ into batches of size $B$) \;
        \For{each local epoch e = 0 to E-1}{
            \For{$b \in \mathcal{B}$}{
                $(\theta, \xi) \leftarrow (\theta, \xi) - \eta \triangledown \mathcal{L}((\theta, \xi); b)$ \;
                $precision_{b} \leftarrow$ (batch precision) \; \label{line:pe-start}
                $precision_{avg} \leftarrow$ (cumulative average precision) \;
                
            }
            \textcolor{comment-color}{// Personalized epoch} \
            
            \If{any $precision_{avg}$ > 0.95 or $precision_b$ == 1}{
                break \; \label{line:pe-end}
            }
        }
        
        \textcolor{comment-color}{// Personalized clustering} \
        
        Merging $m$ clusters, $M \leftarrow M - m$ \; 
        Update pseudo labels $\mathcal{Y}$ with new clusters \;
        Update classifier $\xi$ with new dimension $v \times M$ \;
        
        
        \KwRet $\theta$\;
    }
\end{algorithm}

\subsubsection{\textbf{Personalized Epoch}} 
\label{sec:pe}
We propose personalized epoch (PE) to reassign computations throughout training by dynamically adjusting the number of trained epochs each round. Training with larger local epochs consumes larger computation. 

The majority of federated learning algorithms train for the same local epoch in all rounds \cite{fedavg, zhuang2020fedreid, fedprox}. Researchers design these algorithms for supervised learning. Since data is labeled in supervised learning, using the same computation drives training to achieve better performance, regardless of training rounds. However, data is unlabeled in our scenario, and the pseudo labels are predicted by hierarchical clustering every round. The amount of computation on different rounds would have various impacts on the performance.


Hence, we propose to reassign computation throughout training: using a large local epoch $E$ for the first training round and dynamically adjusting it according to training feedback for the remaining rounds.  We first hypothesize that more computation in the first round is desirable. In the first round, each image is a cluster, which is regarded as a unique identity. Although training with these labels ignores intra-camera and inter-camera relationship of the same identity, it is helpful for learning representations of person ReID images. Compared with labels of the first round, labels predicted in later rounds could be incorrect in visually similar images \cite{lin2019buc}. Since the first training round is not affected by such incorrectness, we propose to train for larger epochs.


For the remaining rounds, we assign enough computation for clients to reach good training precision in each round. As the data statistics of clients and the clustering results of rounds are different, we propose a \textit{Controller} to dynamically adjust the number of local epochs according to real-time training feedback. Specifically, we collect the training precision of each batch and send these feedbacks to the Controller after training of each epoch. The Controller early stops a training round if any of the following conditions are satisfied: (1) the precision of any batch equals 100\%; (2) the cumulative average precision of batches is larger than 95\% (line \ref{line:pe-start}-\ref{line:pe-end} in Algorithm \ref{algo:fedureid}). This early-stop mechanism enables dynamic computation assignment among rounds, even among clients.

\subsubsection{\textbf{Personalized Clustering}} 
\label{sec:pc}
We propose personalized clustering (PC) to enable clients to customize the merge percent $mp$ by approximating the number of identities via profiling. The merge percent $mp$ determines the number of merged clusters in each round ($m = n_k \times mp$), which controls the pace of clustering.

\textit{Clients should not use the same merge percent} because their data vary in the number of images and identities. As the clusters are regarded as pseudo labels (identities), when the number of predicted identities falls below the actual number of identities, some labels are certainly wrongly predicted. It would cause performance drops. Since clients have different numbers of images and identities, they reach the number of clusters below the number of identities at different training rounds. Thus, they suffer from performance drop at different rounds. As such, the model obtained in the server by aggregating clients' models would not be optimal because of the degraded performance of some clients' models.

We propose to enable clients to personalize clustering paces to fit their characteristics of datasets. To mitigate the performance drop caused when the number of clusters is smaller than the actual number of identities $I_k$, a natural idea is to control client $k$ to finish training with $I_k$ clusters. In this way, by fixing the number of training rounds $R$, client $k$ obtains customized merged clusters each round $m = \frac{n_k - I_k}{R}$. However, this solution is not feasible in real-world scenarios --- data is unlabeled, so the actual number of identities is unknown.

To this end, we design a \textit{Profiler} to estimate the number of identities of clients. We profile clients before they start the first training round (line \ref{line:profiling} in Algorithm \ref{algo:fedureid}). Profiling produces the number of merged clusters per round $m_k$ and the merge percent $mp_k$ for client $k$. To minimize the computation overhead caused by profiling, we conduct unsupervised person ReID for each client using larger merge percent $mp_{profile}$ (i.e., fewer rounds) and smaller local epochs $E_{profile}$. After profiling, we analyze the results and choose the round $r_{profile}$ that achieves the best accuracy. For each client, we select the number of clusters $M_{profile}$ in round $r_{profile}$ as the estimated number of identities. As a result, we calculate the merged clusters $m_k$ and the merge percent $mp_k$ for client $k$ with the following formula:
\begin{equation}
    m_k = \frac{n_k - M_{profile}}{R},
\label{eq:pc-1}
\end{equation}
\begin{equation}
    mp_k = \frac{m_k}{n_k},
\label{eq:pc-2}
\end{equation}
where $R$ is the total training rounds and $n_k$ is the data volume of client $k$. The Profiler sends these clustering configurations to the Controller to control the training flow and clustering flow.

\subsection{Server Design}
\label{sec:server-design}

The cloud server is responsible for coordinating clients to conduct training. It aggregates models trained in clients and updates clients with a new aggregated model for the next training round. To address the statistical heterogeneity of clients, we propose an optimization method, \textit{Personalized Update} (PU), to adapt the aggregated model for clients.

\subsubsection{\textbf{Model Aggregation}} At the end of each training round $r$, the server aggregates models uploaded from clients with weighted averages. The weightage of client $k$ depends on its data volume $n_k$. For $K$ participated clients, the model aggregation formula is as follows:
\begin{equation}
    \theta^{r+1} = \sum_{k=1}^K \frac{n_k}{n} \theta^r_k, \;
\label{eq:weighted-average}
\end{equation}
where $n = \sum_{k=1}^K n_k$ is the total data volume. The global model $\theta^{r+1}$ is the generalized model, which is deployable to other scenarios. We also evaluate $\theta^{r+1}$ on test sets of all clients.





\subsubsection{\textbf{Personalized Update}} 
\label{sec:pu}
We propose Personalized Update (PU) to adapt the aggregated model to client $k$ by updating the local model in client $k$ with exponential moving average (EMA) of the global model. The weighting of EMA is measured by the similarity between the global and local two models.

Standard federated algorithms like FedAvg \cite{fedavg} simply updates the models in clients by replacing them with the global model. However, the global model may not fit all clients \cite{fedprox} because of statistical heterogeneity among clients --- the scenes (indoor or outdoor) and illumination could be different. To mitigate statistical heterogeneity, we incorporate the local model $\theta_k^r$ of client $k$ in round $r$ by updating it with an exponential moving average. We formulate the personalized update as follows:
\begin{equation}
    \theta_k^{r+1} = \mu \theta_k^r + (1-\mu) \theta^{r+1},
\label{eq:pu}
\end{equation}
where $\mu$ is the weighting ranging from $[0, 1]$. $\mu$ determines the importance of the local model $\theta_k^r$ and global model $\theta^{r+1} $ in update. Instead of setting $\mu$ as constant in all training rounds in all clients, We propose to calculate $\mu$ based on the similarity of the global and local model (line \ref{line:pu-start}-\ref{line:pu-end} in Algorithm \ref{algo:fedureid}): (1) Calculate the Euclidean distance of each layer of these two models; (2) Normalize distances of layers to [0, 1]; (3) Average these distances as $\mu$. The intuition of calculating $\mu$ is to increase the importance of the local model when the local model is not similar to the global model. As such, the updated model $\theta^{r+1}$ retains more historical information of the local model, which is more personalized to the data in the client.

\begingroup
\setlength{\tabcolsep}{0.4em}
\begin{table}[t]
\caption{The statistics of eight person ReID datasets.}
\begin{tabular}{lccccccc}
\toprule
\multicolumn{1}{l}{\multirow{3}{*}{Datasets}} &
  \multicolumn{2}{c}{Train} &
  \multicolumn{1}{c}{} &
  \multicolumn{2}{c}{Test} \\ 
\cline{2-3} \cline{5-6}
& \multirow{2}{*}{\# IDs} & \multirow{2}{*}{\# Images} & & Query & Gallery \\
& & & & \# Images & \# Images \\
\midrule
DukeMTMC-reID \cite{zheng2017dukemtmc-reid} & 702 & 16,522 & & 2,228 & 17,611 \\
Market-1501 \cite{Zheng2015Market1501} & 751 & 12,936 & & 3,368 & 19,732 \\ 
CUHK03-NP \cite{Li2014CUHK03} & 767 & 7,365 & & 1,400 & 5,332 \\ 
PRID2011 \cite{prid2011} & 285 & 3,744 & & 100 & 649 \\ 
CUHK01 \cite{li2012cuhk01} & 485 & 1,940 & & 972 & 972 \\ 
VIPeR \cite{Gray2008ViewpointIP} & 316 & 632 & & 316 & 316 \\ 
3DPeS \cite{3dpes} & 93 & 450 & & 246 & 316 \\ 
iLIDS-VID \cite{iLIDS-VID} & 59 & 248 & & 98 & 130 \\ 
\bottomrule
\end{tabular}
\label{tab:dataset}
\end{table}
\endgroup

\begin{figure}[t]
\begin{center}
\includegraphics[width=0.9\linewidth]{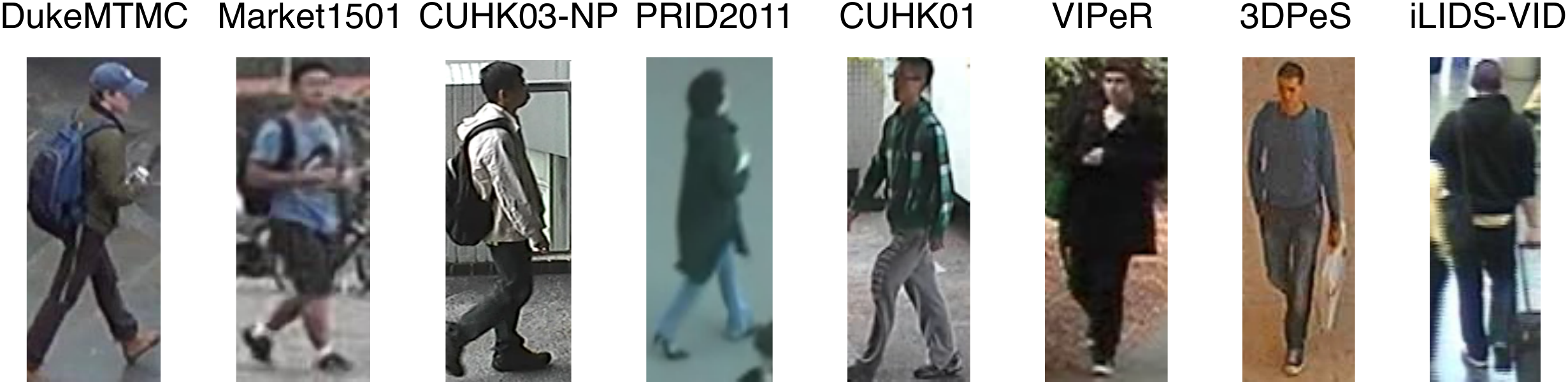}
\end{center}
    \caption{Sample images of eight datasets.}
\label{fig:dataset}
\end{figure}

\begin{table*}[t]
\caption{Performance comparison of FedUReID with the baseline and unsupervised domain adaptation (UDA) methods on the two largest datasets. Despite that FedUReID does not need any labels like UDA methods, it outperforms all other methods by 4\% and 3\% on rank-1 accuracy of DukeMTMC-reID and Market-1501 datasets, respectively.}
\begin{center}
\begin{tabular}{llcccccccccc}
\toprule
\multicolumn{1}{l}{\multirow{2}{*}{Methods}} &
\multicolumn{1}{l}{\multirow{2}{*}{Types}} &
\multicolumn{5}{c}{Market-1501 Dataset \cite{Zheng2015Market1501} (\%)} &
\multicolumn{4}{c}{DukeMTMC-reID Dataset \cite{zheng2017dukemtmc-reid} (\%)}
\\
\cline{3-6} \cline{8-11}
\multicolumn{1}{c}{} &
\multicolumn{1}{c}{} &
\multicolumn{1}{c}{Rank-1} &
\multicolumn{1}{c}{Rank-5} &
\multicolumn{1}{c}{Rank-10} &
\multicolumn{1}{c}{mAP} &
\multicolumn{1}{c}{} &
\multicolumn{1}{c}{Rank-1} &
\multicolumn{1}{c}{Rank-5} &
\multicolumn{1}{c}{Rank-10} &
\multicolumn{1}{c}{mAP}
\\
\midrule
PUL \cite{fan2018unsupervised-tomm-pul} & Domain Adaptation & 44.7 & 59.1 & 65.6 & 20.1 & & 30.4 & 46.4 & 50.7 & 16.4 \\
SPGAN \cite{deng2018spgan} & Domain Adaptation & 58.1 & 76.0 & 82.7 & 26.7 & & 46.9 & 62.6 & 68.5 & 26.4 \\
HHL \cite{zhong2018generalizing-hhl} & Domain Adaptation & 62.2 & 78.8 & 84.0 & 31.4 & & 46.9 & 61.0 & 66.7 & 27.2 \\
BUC \cite{lin2019buc} (Standalone) & Purely Unsupervised & 61.9 & 73.5 & 78.2 & 29.6 & & 40.4 & 52.5 & 58.2 & 22.1 \\
Baseline & Purely Unsupervised & 60.5 & 73.3 & 77.9 & 27.4 & & 47.0 & 58.3 & 64.1 & 25.2 \\
\hline

FedUReID (Ours) & Purely Unsupervised & \textbf{65.2} & \textbf{77.8} & \textbf{82.2} & \textbf{34.2} & & \textbf{51.0} & \textbf{62.4} & \textbf{67.6} & \textbf{29.5} \\ 

\bottomrule
\end{tabular}
\end{center}
\label{tab:comparison}
\end{table*}

\section{Experimental Results}
\label{sec:experiments}

In this section, we start by describing the experimental setup. We then present the overall performance of FedUReID. We end by analyzing the optimization methods via ablation studies.

\subsection{Experimental Setup}

We first present the experimental setups, including datasets, evaluation metrics, implementation details, and experimental settings.

\textbf{Datasets} We evaluate our experiments with eight person ReID datasets adopted from the benchmark, FedReIDBench \cite{zhuang2020fedreid}. Table \ref{tab:dataset} presents the statistics of the datasets. These datasets vary in the number of images and identities. As the datasets are collected from various locations, the visual appearances of them are also quite different, as shown in Figure \ref{fig:dataset}. These characteristics of datasets simulate the statistical heterogeneity in real-world scenarios.

\textbf{Evaluation Metrics} We evaluate the performances with standard ReID evaluation metrics and computation cost. To evaluate the performances of ReID, we use the two most common evaluation metrics: Cumulative Match Characteristic (CMC) curve and mean Average Precision (mAP) \cite{zheng2016person-reid-survey}. Given an image as a query, CMC first ranks gallery images by similarity (from most similar to least similar). It then compares whether the ranked top-k images match the query image. The probability of such matching is denoted as rank-k accuracy. We report the results of rank-1, rank-5, and rank-10 accuracy. We also evaluate the performance with mAP, which measures the mean average precision of all queries.  

We measure the computation cost by the number of local epochs. Although the computation cost of the classifier varies in rounds, it is negligible compared to the computation cost of the ResNet-50 \cite{resnet} backbone. The ResNet-50 costs 2.64 gigaFLOPS (GFLOPS), while the classifier of \textit{max} dimension (16,522) costs only 0.035 GFLOPS, around 1.3\% of the ResNet-50. The classifiers with much smaller dimensions cost even less computation. Therefore, we approximate the computation cost by the number of epochs.


\textbf{Implementation Details} We implement FedUReID in Python using EasyFL \cite{zhuang2021easyfl} based on PyTorch \cite{paszke2019pytorch} framework. The model structure of the backbone is ResNet50 \cite{resnet}. We run experiments with one server and eight clients, where each client trains with one dataset. These datasets are collected from multiple camera views, simulating edges collecting data from cameras, and storing in the Data Store. We run the server on Intel(R) Xeon(R) Gold 6130 CPU and run clients on eight NVIDIA® V100 GPUs, one on each GPU. Model aggregation and model update are conducted through the PyTorch communication backend. For all experiments, we evaluate both local models and the global model in each round. Then, we report the best performance on each dataset among all rounds.


\textbf{Experimental Settings} By default, we use the following experiment settings: batch size $B = 16$, total training round $R = 20$, and merge percent $mp = 0.05$. We set local epoch $E = 5$ for experiments without PE and $E = 20$ for experiments with PE. 

\subsection{Performance Comparison}

We demonstrate the effectiveness of joint optimization by comparing FedUReID with standalone training, the baseline, and several unsupervised domain adaptation (UDA) methods. Standalone training means that each client performs unsupervised person ReID training \cite{lin2019buc} with its dataset --- not collaborating with other clients. It is only meaningful for a client to participate in federated learning (FL) if the performance is better than its standalone training. The baseline method is the simple combination of FL \cite{zhuang2020fedreid} and unsupervised person ReID \cite{lin2019buc}, as described in Section \ref{sec:baseline}.

\begin{figure}[t]
\begin{center}
\includegraphics[width=\linewidth]{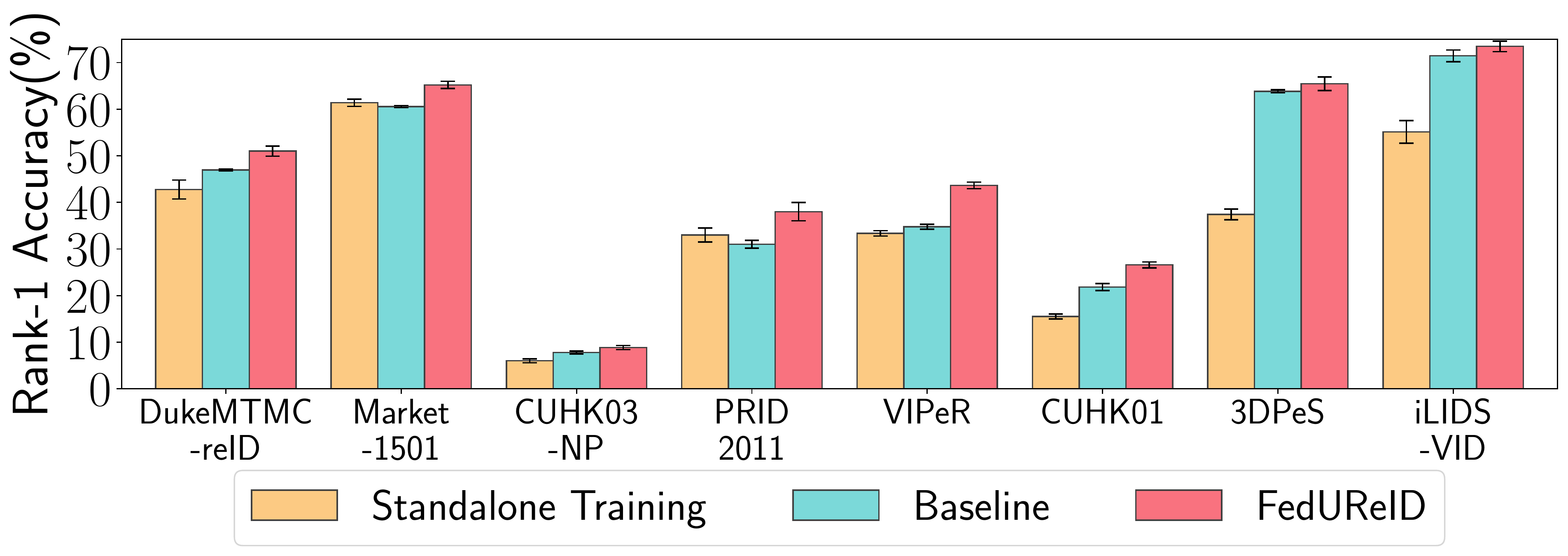}
\end{center}
    \caption{Performance (rank-1 accuracy) comparison of standalone training, the baseline, and FedUReID. FedUReID achieves the best performance on all datasets.}
\label{fig:overall-performance}
\end{figure}

Figure \ref{fig:overall-performance} compares the rank-1 accuracy of FedUReID with standalone training and the baseline. The standalone training is better than the baseline on Market-1501 \cite{Zheng2015Market1501} and PRID2011 \cite{prid2011} datasets. These results indicate that combining FL and unsupervised person ReID is not trivial. It requires deep understanding and analysis to optimize the performance. Our proposed FedUReID, with joint optimization of cloud and edge, outperforms both standalone training and the baseline in all datasets. Compared with standalone training, another insight is that FL-based methods significantly improve performance on smaller datasets (less than 2,000 training images). For example, the improvement is over 26\% on the 3DPeS \cite{3dpes} dataset.

\begingroup
\setlength{\tabcolsep}{0.26em}
\begin{table}[t]
\caption{Ablation studies of our proposed optimization methods: personalized clustering (PC) and personalized epoch (PE) in the edges; personalized update (PU) in the cloud; joint optimization of cloud and edge.}
\begin{center}
\begin{tabular}{lcccccccccc}
\toprule
\multicolumn{1}{l}{\multirow{2}{*}{Datasets}} &
\multicolumn{1}{l}{\multirow{2}{*}{Baseline}} &
\multicolumn{3}{c}{Edge} &
\multicolumn{1}{c}{} &
\multicolumn{1}{c}{Cloud} & 
\multicolumn{1}{c}{} &
\multicolumn{1}{c}{Joint}
\\
\cline{3-5} \cline{7-7} \cline{9-9}
\multicolumn{1}{c}{} &
\multicolumn{1}{c}{} &
\multicolumn{1}{c}{PC} &
\multicolumn{1}{c}{PE} &
\multicolumn{1}{c}{Both} &
\multicolumn{1}{c}{} &
\multicolumn{1}{c}{PU} &
\multicolumn{1}{c}{} &
\multicolumn{1}{c}{All}
\\
\midrule
DukeMTMC-reID\cite{zheng2017dukemtmc-reid} & 47.0 & 48.3 & 49.5 & 50.4 & & 49.2 & & \textbf{51.0} \\
Market-1501\cite{Zheng2015Market1501} & 60.5 & 62.5 & 64.0 & 65.1 & & 62.2 & & \textbf{65.2} \\
CUHK03-NP\cite{Li2014CUHK03} & 7.8 & 8.4 & 7.9 & 8.1 & & 8.8 & & \textbf{8.9} \\
PRID2011\cite{prid2011} & 31.0 & 34.0 & 35.0 & 37.0 & & 36.0 & & \textbf{38.0} \\
CUHK01\cite{li2012cuhk01} & 34.8 & 39.3 & 39.2 & 42.6 & & 35.4 & & \textbf{43.6} \\
VIPeR\cite{Gray2008ViewpointIP} & 21.8 & 24.4 & 24.4 & 24.7 & & 22.5 & & \textbf{26.6} \\
3DPeS\cite{3dpes} & 63.8 & 65.5 & 64.6 & \textbf{67.5} & & 65.0 & & 65.5 \\
iLIDS-VID\cite{iLIDS-VID} & 71.4 & 73.5 & 70.4 & 70.4 & & 72.5 & & \textbf{73.5} \\

\bottomrule
\end{tabular}
\end{center}
\label{tab:ablation}
\end{table}
\endgroup


In addition, Table \ref{tab:comparison} compares FedUReID with three UDA methods: PUL \cite{fan2018unsupervised-tomm-pul}, SPGAN \cite{deng2018spgan}, and HHL \cite{zhong2018generalizing-hhl}, on the two largest datasets. UDA methods improve the performance for unlabeled data in a target domain by heavily relying on plenty of labeled data in a source domain. Despite that FedUReID does not need any labels, it effectively improves the performance by at least 3\% and 4\% on rank-1 accuracy of the Market-1501 \cite{Zheng2015Market1501} and DukeMTMC-reID \cite{zheng2017dukemtmc-reid} datasets, respectively. FedUReID is also superior to other methods on rank-5 accuracy, rank-10 accuracy, and mAP.

\subsection{Ablation Studies}

We conduct ablation studies on the baseline, three proposed optimization methods (PE, PC, and PU), and the combinations of these methods. These ablation studies demonstrate the effectiveness of our proposed optimizations.

We first present the rank-1 accuracy comparison of edge optimizations, the cloud optimization, and joint optimization of cloud and edge. Table \ref{tab:ablation} shows that joint optimization achieves the best overall performance. Also, the client optimization with both PC and PE outperforms either one alone. Besides, any single optimization outperforms the baseline almost on all datasets. Although the performance of PE on iLIDS-VID \cite{iLIDS-VID} is slightly lower than the baseline, it is still much better than standalone training. These results demonstrate that our optimization methods effectively elevate the performance. Next, we analyze the baseline and these optimizations in detail.



\begin{figure}[t]
\begin{center}
\includegraphics[width=\linewidth]{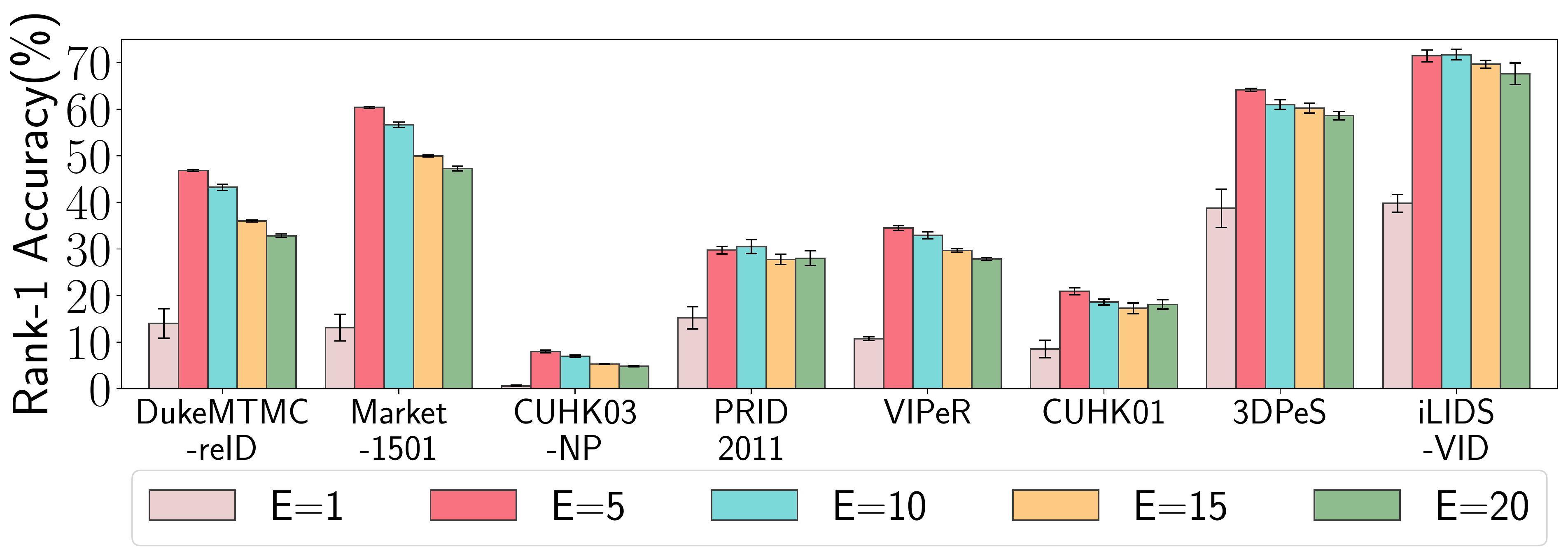}
\end{center}
    \caption{Performance (rank-1 accuracy) comparison of different values of local epochs $E$. We select $E = 5$ as the baseline as it performs better with lower computation costs.}
\label{fig:baseline}
\end{figure}

\textbf{Baseline Method} We select the best setting for the baseline method by comparing performances of different local epochs $E=\{1, 5, 10, 15, 20\}$. Larger local epoch requires higher computation. Following the setting in BUC \cite{lin2019buc}, we fix the merge percent $mp = 0.05$. Figure \ref{fig:baseline} shows that $E = 5$ performs best on rank-1 accuracy. The performance of $E = 10$ is also comparable, but it costs 2x more computation than $E = 5$. Thus, we choose $E = 5$ as our baseline. Besides, the accuracy decreases as increasing computation from $E = 5$ to $E = 20$. This insight indicates that simply increasing computation in all rounds harms the performance. Based on this insight, we propose personalized epoch to reassign computation across different training rounds.

\textbf{Effectiveness of Personalized Epoch} We evaluate personalized epoch (PE) in two folds. Firstly, we evaluate reassigning larger computation to the first training round. For a fair comparison, we maintain the computation similar to the baseline $E = 5$. We train for 20 epochs for the first round and $4 =\lfloor\frac{5 \times 20 - 20}{19}\rfloor$ epochs for the remaining 19 rounds, denoted as \textit{$E=20, 4$}. Secondly, we evaluate dynamically adjusting epochs with an early stop in the remaining rounds, as described in Section \ref{sec:pe}. We use 20 local epochs for the first round and dynamic local epochs for the remaining 19 rounds.

\begingroup
\setlength{\tabcolsep}{0.24em}
\begin{table}[t]
\caption{Performance (rank-1 accuracy) comparison of different local epochs. $E=20, 4$ means training 20 epochs for the first round and 4 epochs for the rest rounds. Personalized epoch (PE) achieves the best overall performance.}
\begin{center}
\begin{tabular}{lcccccccc}
\toprule
Datasets & duke & market & cuhk03 & prid & cuhk01 & viper & 3dpes & ilids
\\
\midrule
Baseline & 47.0 & 60.5 & 7.8 & 31.0 & 34.8 & 21.8 & 63.8 & 71.4 \\
E=20, 4 & 47.9 & 63.6 & 7.4 & 32.0 & 35.4 & 22.8 & \textbf{66.3} & \textbf{71.4} \\ 
PE & \textbf{49.5} & \textbf{64.0} & \textbf{7.9} & \textbf{35.0} & \textbf{39.2} & \textbf{24.4} & 64.6 & 70.4 \\

\bottomrule
\end{tabular}
\end{center}
\label{tab:pe}
\end{table}
\endgroup

\begingroup
\setlength{\tabcolsep}{0.19em}
\begin{table}[t]
\caption{Validation of personalized clustering (PC): (1) The motivation of PC is that clients achieve the best performance at various rounds (Best Round); (2) The profiled numbers of identities are close to the labeled ones; (3) The performance of PC using profiled number of identities is comparable to using the labeled ones.}
\begin{center}
\begin{tabular}{lcccccccc}
\toprule
Datasets & duke & market & cuhk03 & prid & cuhk01 & viper & 3dpes & ilids \\
\midrule
Best Round & 18 & 17 & 14 & 17 & 9 & 0 & 7 & 17 \\
\hline
\multicolumn{9}{l}{\textit{Number of clusters/identities:}} \\
\hline
Labeled & 702 & 751 & 767 & 285 & 485 & 316 & 93 & 59 \\
PC & 670 & 528 & 886 & 156 & 855 & 432 & 90 & 13 \\
\hline
\multicolumn{9}{l}{\textit{Rank-1 accuracy using the number of identities above:}} \\
\hline
Baseline & 47.0 & 60.5 & 7.8 & 31.0 & 34.8 & 21.8 & 63.8 & 71.4 \\
Labeled & 47.7 & \textbf{63.9} & 8.3 & 32.0 & 36.8 & 22.8 & \textbf{66.3} & \textbf{74.5} \\
PC & \textbf{48.3} & 62.5 & \textbf{8.4} & \textbf{34.0} & \textbf{39.3} & \textbf{24.4} & 65.5 & 73.5 \\
\bottomrule
\end{tabular}
\end{center}
\label{tab:pc}
\end{table}
\endgroup

\begin{figure}[t]
    \centering
    \includegraphics[width=.7\linewidth]{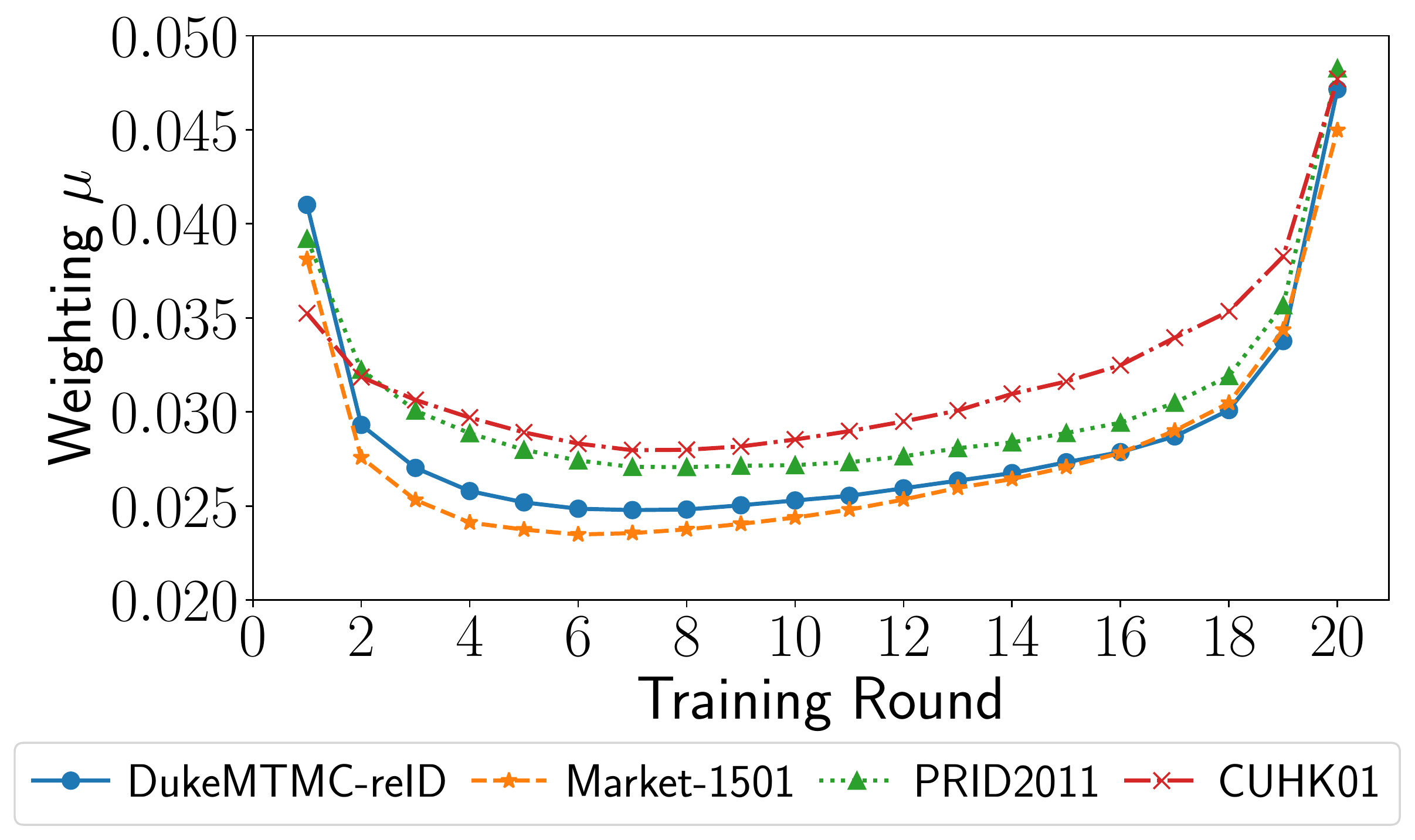}
    \caption{Analysis of personalized update (PU). Measuring the similarity of global and local models with $\mu$. Larger $\mu$ means less similar, i.e. stronger personalization. Models are more personalized at the end of training.}
    \label{fig:similarity}
\end{figure}

Table \ref{tab:pe} compares performances of these three settings of local epochs. A larger epoch on the first round ($E=20, 4$) has better performance than the baseline, except for a small gap on CUHK03 \cite{Li2014CUHK03} dataset. PE (with early stop mechanism) generally outperforms $E=20, 4$. Moreover, PE needs much less computation than $E=20, 4$: $E=20, 4$ needs in total 768 epochs; while PE costs only 479 epochs (Table \ref{tab:computation-comparison}), around 38\% lower than. These results indicate that PE achieves better performance with lower computation costs.

\textbf{Effectiveness of Personalized Clustering} We evaluate personalized clustering (PC) in three folds. Firstly, we validate our motivation that clients should personalize clustering paces. \textit{Best round} in Table \ref{tab:pc} represents the round that \textit{standalone training} achieves the best performance when the merge percent is the same for all clients. The best round of clients are different, so we propose PC to customize their merge percent. Secondly, we present that the profiled number of identities is similar to the actual \textit{labeled} ones (Table \ref{tab:pc}). To minimize computation and ensure performance at the same time, we profile with using following settings: merge percent $mp_{profile} = 0.08$, total rounds $r_{profile} = 12$, and 5 epochs for the first round and 1 epoch for the remaining rounds $E_{profile} = 5, 1$. Profiling incurs 16 epochs of extra computation in each client, 128 epochs in total. Thirdly, Table \ref{tab:pc} also shows that PC (using the profiled number of identities) is comparable to using the actual labeled number of identities, both outperforming the baseline. It demonstrates the effectiveness of profiling and PC.

\begin{figure}[t]
    \begin{subfigure}{.63\linewidth}
        \centering
        \includegraphics[width=\linewidth]{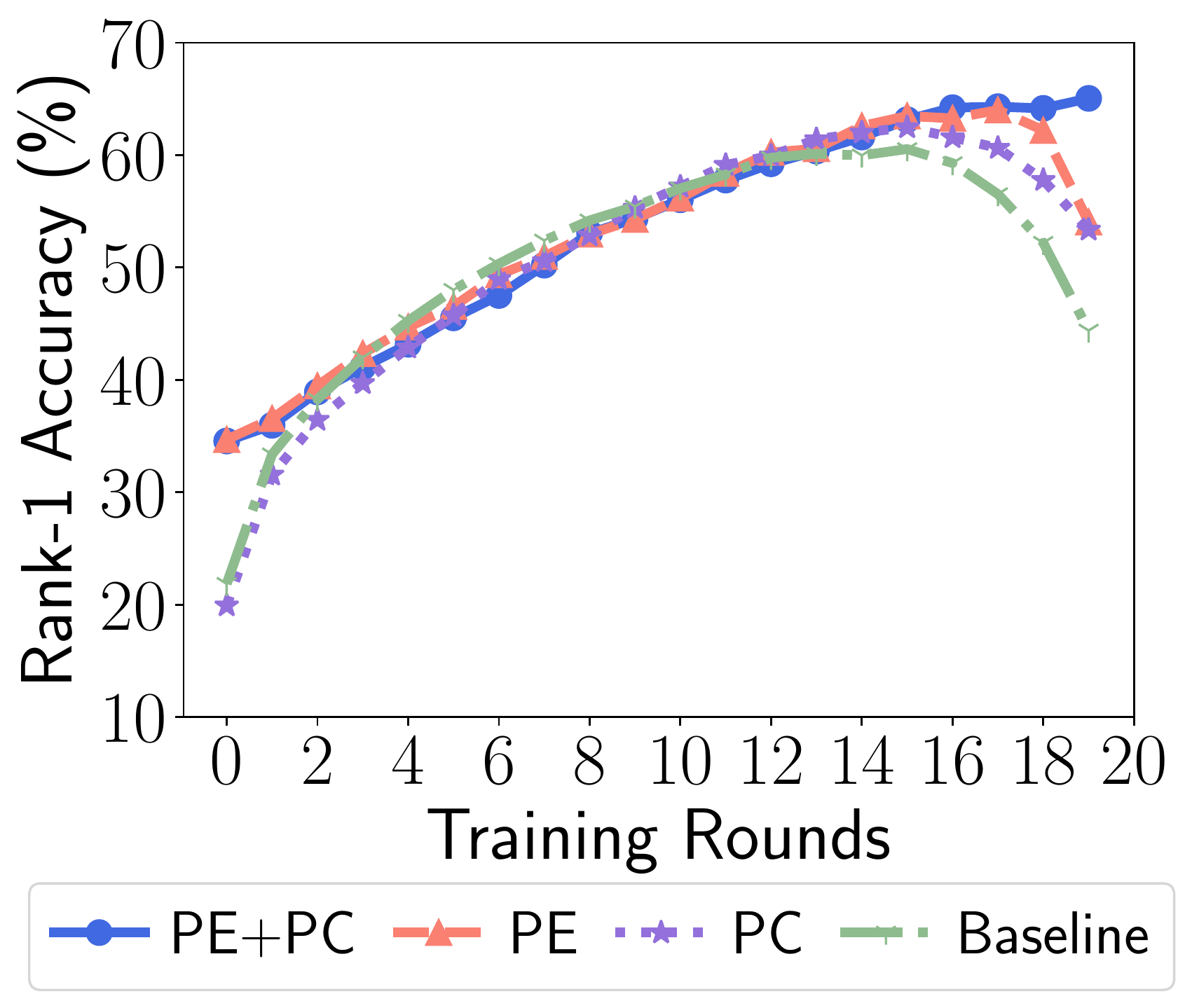}
        \caption{Market-1501 (Large Dataset)}
    \label{fig:pc-pe-a}
    \end{subfigure}%
    \hfill
    \begin{subfigure}{.63\linewidth}
        \centering
        \includegraphics[width=\linewidth]{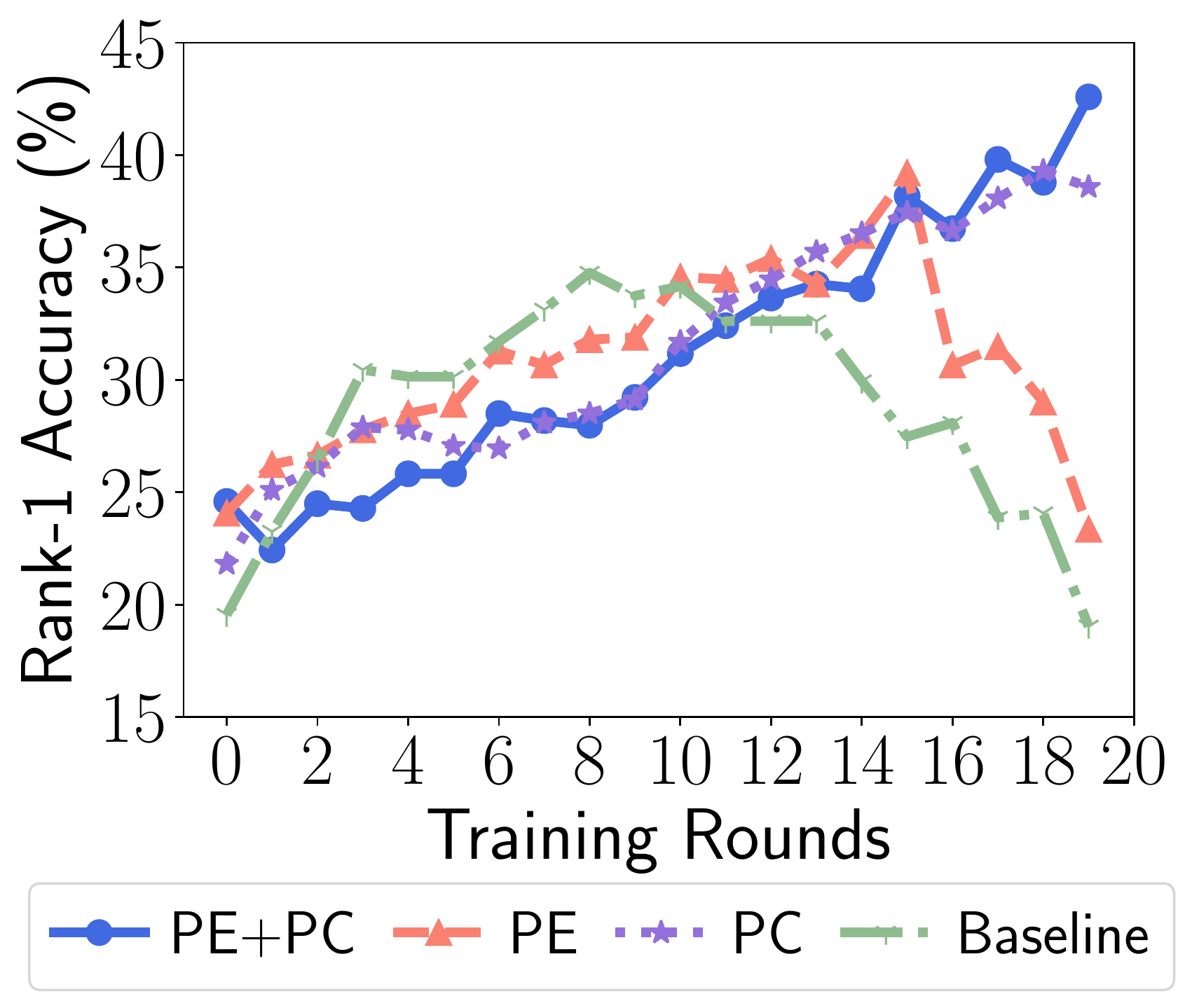}
        \caption{CUHK01 (Small Dataset)}
        \label{fig:pc-pe-b}
    \end{subfigure}

    \caption{Performance comparison throughout training of combined optimization (PE+PC), PE, PC, and the baseline. Combined client optimization retains respective advantages of PE and PC on both large and small datasets, increasing performance as training proceeds.}
    
    \label{fig:pc-pe}
\end{figure}

\textbf{Effectiveness of Personalized Update} we analyze personalized update (PU) by understanding changes in similarities between the global model and local models. The similarity is measured by $\mu$ as described in Section \ref{sec:pu}. Two models are less similar if the value of $\mu$ is larger, so \textit{larger $\mu$ means stronger personalization}. Figure \ref{fig:similarity} shows the changes of similarity of four datasets throughout the training process. All these datasets share the same trend in two stages: (1) $\mu$ decreases in the first few training rounds --- the similarity increases as clients are learning knowledge from the others; (2) $\mu$ increases in the second stage --- the similarity decreases as clients personalize models for their local datasets. Models are more personalized at the end of training. It validates the effectiveness of PU to personalize the global model to models in edges. Besides, PU effectively improves the performance, as compared in Table \ref{tab:ablation}. We provide the results of other datasets in the supplementary.

\begingroup
\setlength{\tabcolsep}{0.36em}
\begin{table}[t]
\caption{Computation cost comparison of different methods. Compared with the baseline, personalized epoch (PE) effectively reduces computation cost by 40\%. The joint optimization reduces computation cost by 29\%.}
\begin{center}
\begin{tabular}{l|c|c|c|c|c}
Methods & Baseline & PE & PC & PE+PC & Joint \\
\hline
Computation Cost (Epochs) & 800 & 479 & 928 & 570 & 566 \\

\end{tabular}
\end{center}
\label{tab:computation-comparison}
\end{table}
\endgroup

\textbf{Effectiveness of Client Optimization} The combined client optimization of PE and PC retains the advantage of both methods. Figure \ref{fig:pc-pe} shows the changes in performances as training proceeds on Market-1501 \cite{Zheng2015Market1501} and CUHK01 \cite{li2012cuhk01} datasets. Market-1501 contains 12,936 training images, representing the larger datasets among eight datasets. While CUHK01 contains 1,940 training images, representing the smaller datasets. PE performs better in larger datasets like Market-1501 (Figure \ref{fig:pc-pe-a}), while PC performs better in smaller datasets like CUHK01 (Figure \ref{fig:pc-pe-b}). Client optimization retains respective advantages and achieves better performances in all datasets. Furthermore, these figures validate our hypothesis that the performance of the baseline drops in the last few rounds. PC and PE both defer and reduce such degradation. Their combination is free from performance drop, achieving the best performance in the last round. We provide figures of other datasets in the supplementary. 

\textbf{Computation Cost} Table \ref{tab:computation-comparison} compares the computation cost of several methods. Compared with the baseline, PE reduces computation cost by 40\% via the early stop of training. PC increases computation cost by 16\% because of profiling. The joint optimization mitigates extra computation costs from PC, reducing 29\% of computation cost, at the same time, achieving the best performance.

\section{Conclusion}
\label{sec:conclusion}

In this paper, we present \textit{FedUReID}, a new federated unsupervised person ReID system to learn models without any labels while preserving privacy. To address the statistical heterogeneity among edges, we propose joint optimization of cloud and edge to personalize models for each edge. For optimizations in edge, we design a Controller to support personalized epoch and a Profiler to facilitate personalized clustering. For the optimization in the cloud, we introduce personalized update to adapt the cloud aggregated models to edges. Extensive empirical studies demonstrate that FedUReID effectively elevates performance on all datasets and reduces computation cost by 29\%.  For future work, we will consider the system heterogeneity among edges.

\begin{acks}
This study is supported by 1) supported under the RIE2020 Industry Alignment Fund – Industry Collaboration Projects (IAF-ICP) Funding Initiative, as well as cash and in-kind contribution from the industry partner(s); 2) the National Research Foundation, Singapore, and the Energy Market Authority, under its Energy Programme (EP Award <NRF2017EWT-EP003-023>); 3) Singapore MOE under its Tier 1 grant call, Reference number RG96/20.
\end{acks}

\bibliographystyle{ACM-Reference-Format}
\bibliography{references}

\newpage


\end{document}